\documentclass[runningheads]{llncs}
\usepackage{subcaption}
\usepackage[utf8]{inputenc}
\usepackage{hyperref}
\usepackage{url}
\usepackage[neverdecrease]{paralist}
\usepackage{graphicx}
\usepackage{booktabs}
\usepackage{graphics}
\usepackage{tabularx}
\usepackage{xspace} 
\usepackage{graphicx}
\usepackage{graphicx} 
\usepackage{booktabs}
\usepackage{mdwlist}
\usepackage{multirow}
\usepackage{amsmath}
\usepackage{amssymb}
\usepackage{enumitem} 
\usepackage{calc}
\usepackage{colortbl}
\usepackage[neverdecrease]{paralist}
\usepackage[ruled,linesnumbered]{algorithm2e}
\usepackage{booktabs}
\usepackage{mathtools}
\usepackage{pifont}
\usepackage{color,soul}
\usepackage{array}
\usepackage{tcolorbox}
\usepackage{color}
\usepackage{framed}
\usepackage{xcolor, soul}
\sethlcolor{yellow}
\usepackage{multirow}

\usepackage[noend]{algpseudocode}
\usepackage{adjustbox}

\usepackage{balance}

\usepackage[show]{chato-notes}

\newcommand{\para}[1]{\paragraph{\textnormal{\textbf{#1.}}}} 

\usepackage{tikz}
\newcommand\mybox[2][]{\tikz[overlay]\node[fill=blue!20,inner sep=2pt, anchor=text, rectangle, rounded corners=1mm,#1] {#2};\phantom{#2}}

\newcommand{\monoelectra}{MonoElectra}

\newcommand{\uls}{\begin{itemize}[leftmargin=*]}
\newcommand{\ule}{\end{itemize}}
\newcommand{\ols}{\begin{enumerate}[leftmargin=*]}
\newcommand{\ole}{\end{enumerate}}
\newcommand{\li}{\item}

\newcommand\MyBox[2]{
  \fbox{\lower0.75cm
    \vbox to 1.7cm{\vfil
      \hbox to 1.7cm{\hfil\parbox{1.4cm}{#1\\#2}\hfil}
      \vfil}%
  }%
}

\title{
Evaluating the Explainability of Neural Rankers}

\author{
Saran Pandian\inst{1} \thanks{The work was conducted at Dhirubhai Ambani Institute of Information and Communication Technology, the author's previous affiliation.} \and
Debasis Ganguly\inst{2} \and
Sean MacAvaney\inst{2}
}

\authorrunning{Pandian et al.}

\institute{
University of Illinois, Chiacgo, USA
\and
University of Glasgow, Glasgow, UK
\\
\email{spand43@uic.edu},
\email{Debasis.Ganguly@glasgow.ac.uk},
\email{Sean.Macavaney@glasgow.ac.uk}
}

\begin{document}

\maketitle

\begin{abstract}

Information retrieval models have witnessed a paradigm shift from unsupervised statistical approaches to feature-based supervised approaches to completely data-driven ones that make use of the pre-training of large language models. While the increasing complexity of the search models have been able to demonstrate improvements in effectiveness (measured in terms of relevance of top-retrieved results), a question worthy of a thorough inspection is - ``how explainable are these models?'', which is what this paper aims to evaluate.
In particular, we propose a common evaluation platform to systematically evaluate the explainability of any ranking model (the explanation algorithm being identical for all the models that are to be evaluated). In our proposed framework, each model, in addition to returning a ranked list of documents, also requires to return a list of explanation units or rationales for each document. This meta-information from each document is then used to measure how locally consistent these rationales are as an intrinsic measure of interpretability - one that does not require manual relevance assessments. Additionally, as an extrinsic measure, we compute how relevant these rationales are by leveraging sub-document level relevance assessments.
Our findings show a number of interesting observations, such as sentence-level rationales are more consistent, an increase in complexity mostly leads to less consistent explanations, and that interpretability measures offer a complementary dimension of evaluation of IR systems because consistency is not well-correlated with nDCG at top ranks. 

\end{abstract}


\section{Introduction}

A neural ranking model (NRM) involves learning a data-driven parameterised similarity function between queries and documents~\cite{colbert,monot5,ance,deepct,nogueira2019doc2query}. Despite achieving state-of-the-art effectiveness (measured in terms of relevance of search results), NRMs suffer from poor interpretability of their underlying working mechanism due to two main reasons. First, they lack a closed form expression involving the human interpretable fundamental components of an IR similarity function (i.e., term frequency, inverse document frequency and document length)~\cite{DBLP:conf/sigir/SenGVJ20}. Second, due to the fact that the similarity function of an NRM operates at the level of embedded representations of documents and queries, it is difficult to determine which terms present within a document are mainly responsible for contributing to its retrieval status value~\cite{explainable-ir-survey}.
To the best of our knowledge, there is no work in IR research that has evaluated the quality of explanations of NRMs in an objective manner, e.g., in terms of consistency and correctness akin to some of the work done for the broader class of predictive models \cite{oramas2018visual,data_repr_GTE,bim}. 



\para{Our Contributions}
First, we introduce a framework for an offline evaluation of the explainability of a ranking model. The evaluation protocol requires participating IR systems to report a list of explanation units or rationales comprised of text snippets of arbitrary lengths in addition to a ranked list of retrieved documents.
Second, we show how these rationales can be evaluated both in an intrinsic and extrinsic manner - the later requiring sub-document level relevance.
We conduct a range of experiments to compare the explanation qualities across a range of different NRMs using a common explanation methodology - that of occlusion commonly used in the literature \cite{li2016understanding,lime,SHAP}.
An important finding of our experiments is that we demonstrate that the IR systems that produce the most relevant results are not necessarily the most explainable.

\section{Related Work} \label{sec:RelWork}

In IR, there has been little effort in investigating the explanation effectiveness of neural models. The article \cite{verma2019lirme} proposes a LIME-based local explanation methodology for IR models and also proposes a metric that measures the overlap of explanation units with the relevant terms (bag-of-words representation of the set of relevant documents for a query). However, the metric proposed in this paper is different from that of \cite{verma2019lirme} in that our metric factors in intrinsic consistency of the explanations, and our proposed extrinsic measure makes use of sub-document level relevance.

Another thread of work towards a quantitative evaluation of IR model explanations is in the form of how well a simple linear model either comprised of the fundamental functional components - term frequency, idf, document length etc. fits a complex black-box model \cite{DBLP:conf/sigir/SenGVJ20}, or that how well it conforms to the axioms of IR \cite{DBLP:conf/ecir/LyuA23} yielding a notion of fidelity. An yet another approach of IR trustworthiness evaluation involves the notion of measuring how consistently does an IR model conform to user expectations of the similarity of the top-retrieved documents based on information need change across query reformulations \cite{Sen0GVR22}.

Our evaluation framework for IR explanation effectiveness is different from those of \cite{DBLP:conf/sigir/SenGVJ20,DBLP:conf/ecir/LyuA23} in the sense that ours does not involve evaluating how well a simpler model fits a more complex one, and it also differs from \cite{Sen0GVR22} in that our evaluation directly concerns model explanations in the form of rationales instead of analysing a model's behaviour across pairs of queries.

\section{Evaluation Framework} \label{sec:evalframework}

\begin{figure}[t]
\centering
\includegraphics[width=.7\columnwidth]{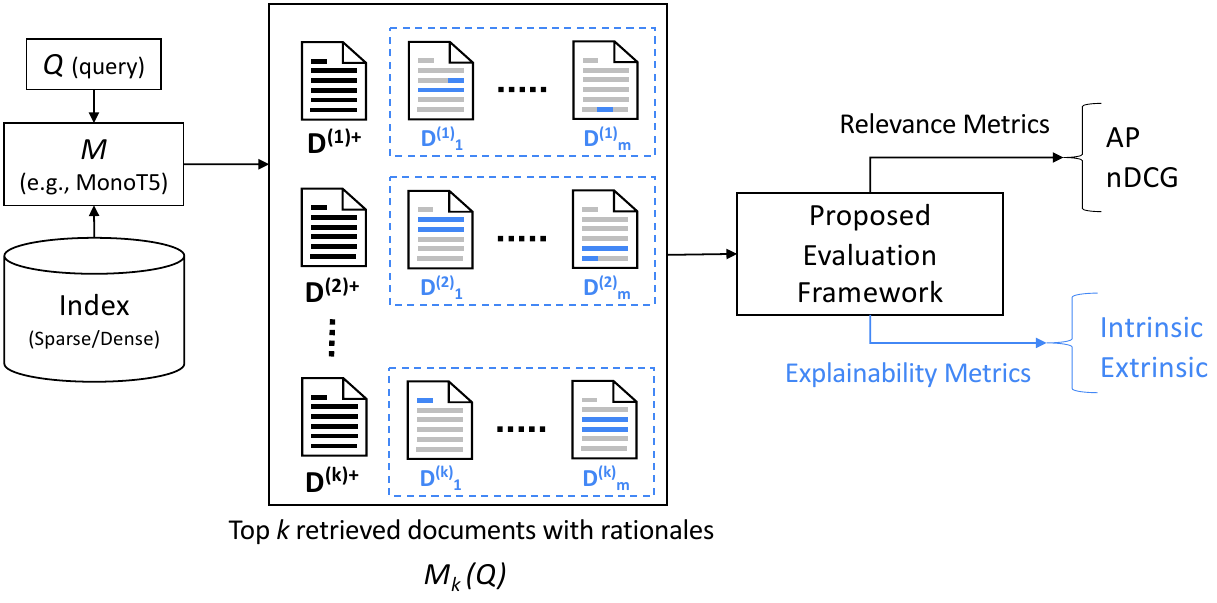}    
\caption{
\small
The proposed workflow for measuring effectiveness of explainable IR models via intrinsic and extrinsic explanation effectiveness measures in addition to relevance-based ones. The meta-information comprised of the explanation units or rationales (shown in blue) is the additional output either obtained from a ranking model itself or with the help of a common explanation methodology (as is the case in our experiments). 
}
\label{fig:schematic_explanation}
\end{figure}

In our proposed evaluation workflow, each IR model $M^\theta$, in addition to outputting a ranked list of $k$ documents $M^\theta_k(Q)$ for a query $Q$, also outputs a list of the most important text segments as \textit{rationales} for each retrieved document. Our evaluation is oblivious of the exact method by which these explanations are obtained for each document - possible options being applying a local explanation algorithm for post-hoc per-document explanations via LIME \cite{lime,verma2019lirme}, SHAP \cite{SHAP} etc., or by making this process an integral part of an IR model by application of feature occlusion based methods \cite{li2016understanding,ZeilerF14} to estimate the importance at sub-document level.
%

\para{Intrinsic Explainability Measure}
Given a list of $m$ rationales for each top-ranked document $D_i$ ($D_i \in M^\theta_k(Q)$, i.e., the top-$k$ set for a query $Q$), we obtain its representation $D^{(i)}$ by concatenating the text from each rationale (each being a segment of $D_i$ itself), i.e.,
\begin{equation}
\phi_m(D) = D^{(i)} = d^{(i)}_1 \oplus \ldots d^{(i)}_m.
\label{eq:pseudodoc}
\end{equation}

Next, to measure how \textbf{consistent} are these rationales, we use this rationale-based representation of each document to recompute its score with the black-box neural model. Subsequently, we rerank the top-$k$ with these modified scores and compute the agreement of this re-ranked top-$k$ with the original one, i.e., the ranking agreement between the explanation-based representation of documents and their original ones.
Formally speaking, we define the consistency metric as Mean Rank Correlation (MRC) as
\begin{equation}
\text{MRC}(\theta) = \frac{1}{|\mathcal{Q}|}\sum_{Q \in \mathcal{Q}}\sigma(M^\theta_k(Q), \hat{M}^\theta_k(Q)), 
\label{eq:mrc}
\end{equation}
where $\mathcal{Q}$ is a set of queries, $M^\theta_k(Q)$ denotes the top-$k$ set obtained with the original document content, $\hat{M}^\theta_k(Q)$ denotes the re-ranked set of documents scored with the meta-information of explanation rationales obtained with $\phi$ (i.e., the representation $D^{(i)}$ of Equation \ref{eq:pseudodoc}), and $\sigma$ denotes a rank correlation measure, such as Kendall's $\tau$.   

The reason the metric MRC (Equation \ref{eq:mrc}) addresses the consistency of an explanation model can be attributed to the following argument. Consider two neural models $\theta_1$ and $\theta_2$ both with the same underlying explanation mechanism $\phi$, which is either a part of the model or a stand-alone local explanation methodology such as LIME for ranking \cite{verma2019lirme}. A higher value of $\text{MRC}(\theta_1)$ indicates that the model $\theta_1$ provides more \textit{faithful explanations} for its observed top list of documents as compared to $\theta_2$, because the model $\theta_1$ when presented with only the rationales for the top documents still scores them in a relatively similar manner thus ensuring a higher correlation with the original list. The value being lower for $\theta_2$, on the other hand, indicates that the rationales themselves do not reflect the true reason behind the relative score computation of the model $\theta_2$.

\para{Extrinsic Explainability Measure} 

The evaluation metric MRC of Equation \ref{eq:mrc} is intrinsic in nature because it does not rely on the availability of relevance assessments of the rationales themselves. Consequently, while MRC is useful to find if a model is more consistent in its explanations than another, it cannot explicitly answer if a model's rationales align well with a human's perception of relevance.
%
Assuming that each document $D_i \in M^\theta_k(Q)$ is comprised of a list of $n_i$ relevant passages (may be sentences or paragraphs) of the form $R(D_i)=\{r^{(i)}_1,\ldots,r^{(i)}_{n_i}\}$\footnote{
If $D_i$ is non-relevant, $n_i=0$ and $R(D_i) = \emptyset$.},
the main idea now is to check whether the rationales overlap with the relevant passages. Since the explanation rationales are arbitrary segments of text without being restricted to passage boundaries, a simple and effective way of measuring the degree of overlap between the explanation units and the relevant passages is to aggregate the matches in their content. More precisely, we compute MER (Mean Explanation Relevance) as
\begin{equation}
\text{MER}(\theta) = \frac{1}{|\mathcal{Q}|mk}
\sum_{Q \in \mathcal{Q}}
\sum_{i=1}^k
\sum_{j=1}^{m} \max_{j'=1}^{n_i} \omega(d^{(i)}_j, r^{(i)}_{j'}), \label{eq:mer} 
\end{equation}
where $\omega$ is a similarity function. We use the cosine-similarity as a concrete realisation of $\omega$.
%
Similar to MRC, a high value of the MRE metric is preferable because it indicates that the rationales demonstrate a substantial overlap with relevant content\footnote{Implementation of the proposed explainability evaluation measures are available at \url{https://github.com/saranpandian/XAIR-evaluation-metric}}.

Note that in our evaluation setup, we use \textbf{the same evaluation algorithm for all IR systems}, which means that our metrics \textbf{do not report how effective is an explanation model} itself (akin to, e.g., the prior work of comparing the fidelity scores of LIME with SHAP \cite{SHAP}); rather these metrics in our setup capture \textbf{how effectively can an IR model be explained} with a specific explanation algorithm.

\section{Experiment Setup}


\subsection{Research Questions, Datasets and IR Models}
We investigate the following research questions in our experiments.

\uls
\li \textbf{RQ-1}: What are the relative variations in the explanation consistency (intrinsic evaluation measure - MRC) across different IR models, i.e., are some models more \emph{explainable} than others?
\li \textbf{RQ-2}: Does MRC provide an aspect of system evaluation that is complementary to that of relevance?
\li \textbf{RQ-3}: What are the relative variations in the relevance-based explanation consistency (i.e, the extrinsic evaluation measure - MER) across different IR models?
\li \textbf{RQ-4}: Does the extrinsic explanation evaluation measure (MER of Equation \ref{eq:mer}) also induce a different relative ordering of IR systems as compared to evaluating them only by relevance?
\ule

For investigating the first two research questions, we conducted experiments on the passage and document collections of MS-MARCO \cite{msmarco-data}.
As topic sets for our experiments, we use the TREC DL'20 topic set, comprising 54 queries.
The reason we employed the two different collections - one where the retrievable units are short passages (average $3.4$ sentences for the MS-MARCO passages), and the other where they are much larger (average of $55.7$ sentences for MS-MARCO documents) is to investigate the effect of document length on the intrinsic and the extrinsic explainability measures.

Recall that computing the extrinsic explainability measure of Equation \ref{eq:mer} requires the availability of sub-document level information. Although the MS-MARCO document collection is constituted of the text units of the MS-MARCO passage collection, they use different identifiers, as a result of which it is not possible to directly use the relevance assessment data of the passage collection as the desired sub-document level relevance information of the document collection. To construct an approximate sub-document level assessments of the MS-MARCO document collection, we make use of passage-level relevance in the document corpus using a document-passage mapping. We built this mapping by matching through the ``QnA'' version of the MS-MARCO dataset\footnote{\url{https://github.com/microsoft/MSMARCO-Question-Answering}}, which provides the URLs of each passage. Documents were matched to the MS-MARCO document ranking corpus IDs via URL, and passages were matched to the MS-MARCO passage ranking corpus IDs through exact text matching (after correcting for character encoding errors present in the passage corpus but not in QnA).
%

We assess the explainability of a variety of neural ranking models (NRMs) with a diverse set of architectures and training regimes in this work.

\uls
\li \textbf{BM25}~\cite{Okapi} is a classic lexical retrieval model with a closed form functional expression involving term frequency, inverse document frequency and document length.
We include BM25 in our experiments mainly as a point of comparison for the explanation effectiveness results of the NRMs. 


\li \textbf{ColBERT}~\cite{colbert} is a multi-representation, `late interation' model that computes relevance based on the sum of the maximum query token similarities in a given document. It represents a strong late interaction model.

\li \textbf{TCT-ColBERT}~\cite{lin-etal-2021-batch} is a single-representation dense retrieval model that was trained from ColBERT using distillation\footnote{We use the \texttt{castorini/tct\_colbert-msmarco} model checkpoint.}. It represents a strong bi-encoder dense retrieval model.


\li \textbf{MonoT5}~\cite{nogueira-etal-2020-document} is a cross-encoder model that is trained to predict ``true'' or ``false'' given a prompt that includes the query and document text. We use the `castorini/monot5-base-msmarco' model checkpoint.

\li \textbf{\monoelectra}~\cite{10.1007/978-3-030-99736-6_44} is a cross-encoder model based on the ELECTRA foundation trained with hard negatives~\cite{Clark2020ELECTRAPT}\footnote{We use the \texttt{crystina-z/monoELECTRA\_LCE\_nneg31} model checkpoint.}.
\ule

All the NRMs in our experiments operate with the re-ranking based setting, i.e., they employ a sparse index (specifically BM25 to retrieve the top-1000 results) which are then re-ranked by the NRM. 
Following the reranking step, we generate the meta-information in the form of rationales for the top $k=10$ and $k=50$ top documents (more details in Section \ref{ss:occlusion}). 

To obtain the top-$k$ for the document ranking task, we first segmented each document into non-overlapping chunks of $3$ sentences, which ensures that the content fits within the $512$ token limit of the underlying transformer models of the NRMs. Next, the score of a document is obtained by taking an aggregation over the scores of the individual chunk. Specifically, we used `$\text{max}$' as an aggregation operator to compute the score of a document as reported in \cite{zhang2020evaluating}.

\subsection{Explanation Model Settings}  \label{ss:occlusion}

As the explanation model for
generating the rationales in our proposed evaluation framework (the meta-information corresponding to each document retrieved in the top-list),
we use an occlusion-based approach \cite{li2016understanding}.
The main advantage of an occlusion-based approach (commonly as counter-factual explanations in recommender models \cite{tan2021counterfactual}) is its run-time efficiency in comparison to approaches such as LIME for ranking (LIRME \cite{verma2019lirme}) which fit a linear regressor to the data of relative score changes with occlusion. We note that the occlusion-based explanations are \textit{not tied to a specific model architecture}; they can be applied to any relevance model that operates on the text of the query and document.

\para{Explanations for Passage Retrieval}
We employ two different granularities for generating the rationales - the first at the level of sentences, and the second at the level of word windows of length $w$ ($w$ being a parameter). Accordingly, we sample $n$ segments (i.e., $n$ number of sentence or word windows as per the chosen granularity, $n$ being a parameter) from a document $D$, and then mask out the selected segments to construct a pseudo-document $D'$.
%
%
Similar to \cite{verma2019lirme}, we then compute the relative score change induced by this masking process, which as per the local explanation principle yields the relative importance of the masked segment.
The importance weights across the $n$ segments are then distributed uniformly.
The occlusion-based weights are then accumulated for each segment after each sampling step.
Formally,
\begin{equation}
\phi(d_i) = \frac{1}{n}\frac{|\theta(Q, D) - \theta(Q, D - \bigcup_{j=1}^nd_j)|}{\theta(Q, D)}  \label{eq:rationale_wt}  
\end{equation}
where $d_i$ a segment of $D$ ($i=1,\ldots,n$).
Finally, we output the top-$m$ segments with the highest weights (as per the $\phi$ values computed via Equation \ref{eq:rationale_wt}) as the rationales for retrieving $D$ for query $Q$.

\para{Explanations for Document Retrieval}

The process is largely  similar to that of passage retrieval with a small number of differences as follows. Firstly, due to the large size of the documents ($55.7$ sentences on an average), it is not possible to encode an entire document's representation via a transformer model during the training or the inference phase. Due to this reason, we partition each document into fixed length chunks, and then obtain the overall score of the document by computing the maximum over the individual scores for each chunk \cite{zhang2020evaluating}. As the number of sentences defining a chunk, we use the value of $3$ (this was chosen so that each chunk in the document collection is approximately of the same size as the retrievable units of the MS-MARCO passage collection, the average length being $3.4$ sentences).

Secondly, due to the large length of the documents we restrict the minimum granularity of rationales to individual sentences. Similar to explaining passage retrieval with $m$ word windows, for document retrieval experiments also we report $m$ top explanation units with the highest fidelity scores (similar to Equation \ref{eq:rationale_wt}), the explanation units being sentences for document retrieval.
We use the same parameter name - $m$ to indicate the number of explanations units; whether this applies for a word or a sentence is to be understood from the context.
%
We iterate through each sentence in a chunk and compute the relative score change induced by its occlusion, i.e.,
\begin{equation}
\phi(d_i) = \frac{\theta(Q,D) - \theta(Q,D-d_i)}{\theta(Q,D)}.
\end{equation}
We then employ a greedy approach and select the one with the highest $\phi(d_i)$ value and remove it from the document $D$. We then the repeat this step $m-1$ more times to eventually yield a total of $m$ sentences as the rationales for an IR model $\theta$.

\section{Explainability Evaluation}


\begin{table}[t]
\centering
\caption{
\small
A comparison of the relevance and the explanation consistency of different IR models on the MS-MARCO passage collection for top-10 and top-50 search results with both sentences and word windows employed as explanation units. For these experiments, we set the size of the word windows to $5$ and the number of rationales to $6$ for the word windows, and $1$ for the sentences.
The best results in terms of relevance and the explanation effectiveness of the NRMs are bold-faced.
\label{tab:passageres}
}
\small
\begin{adjustbox}{width=0.7\columnwidth}
\begin{tabular}{@{}l cc cc cc@{}}
\toprule
& \multicolumn{2}{c}{} & \multicolumn{4}{c}{Explanation Effectiveness} \\
\cmidrule(r){4-7}
 & \multicolumn{2}{c}{Relevance (nDCG)} & \multicolumn{2}{c}
{MRC (Sentences)} & \multicolumn{2}{c}{MRC (Word windows)} \\
\cmidrule(r){2-3}
\cmidrule(r){4-5}
\cmidrule(r){6-7}
Model & top-10 & top-50 & top-10 & top-50 & top-10 & top-50 \\ 
\midrule
BM25 & \cellcolor{lightgray}0.4910 & \cellcolor{lightgray}0.4889 & \cellcolor{lightgray}0.4000 & \cellcolor{lightgray}0.1503 & \cellcolor{lightgray}0.3029 & \cellcolor{lightgray}0.1232\\
\midrule
ColBERT & 0.6900 & 0.6400 & \textbf{0.4502} & 0.1923 & \textbf{0.2790} & 0.1058\\
TCT-ColBERT & 0.6900 & 0.6328 & 0.2938 & 0.1480 & 0.1926 & 0.0974\\
MonoT5 & 0.7133 & 0.6660 & 0.3481 & 0.2230 & 0.1844 & 0.1102\\
MonoElectra & \textbf{0.7460} & \textbf{0.6958} & 0.4165 & \textbf{0.2262} & 0.2644 & \textbf{0.1303}\\

\bottomrule
\end{tabular}
\end{adjustbox}
\end{table}

\para{Passage Ranking Results}
Table \ref{tab:passageres} presents the results of evaluating different IR systems on the MS-MARCO passage collection in terms of relevance and our proposed intrinsic explanation evaluation measure MRC. The extrinsic explanation measure - MER cannot be computed for the MS-MARCO passage collection as there is no sub-document level information available for these retrievable units.
We now present the following key observations from Table \ref{tab:passageres}.

\textbf{Relevance-based observations:}
As expected, all the NRMs significantly outperform BM25 in terms of relevance-only based evaluation, which is consistent with the findings of existing research \cite{craswell2020overview}.

\textbf{Relevance vs. Explainability:}
Interestingly, the model with the best retrieval effectiveness is not necessarily the most well-explainable model.
This can be seen from the fact that the MRC@10 values for the best performing model in terms of relevance, i.e., \monoelectra, are not the best ones (ColBERT MRC@10 values are better).   
This answers RQ-2 in affirmative in the sense that the explanation quality potentially provides an alternate way towards IR model evaluation different from that of relevance.

\textbf{NRMs with the best explanation consistencies:}
BM25 mostly yields more consistent explanations in comparison to NRMs. This can be seen by comparing the gray values under the two column groups with the corresponding non-gray ones across the same column, e.g., BM25 achieves the highest MRC@10 of $0.3029$ with word-window based rationales, which is higher than the best explanation effectiveness obtained by an NRM ($0.2790$ with ColBERT).  

\begin{figure}[t]
\centering    
\begin{subfigure}[b]{0.32\textwidth}
    \includegraphics[width=\textwidth]{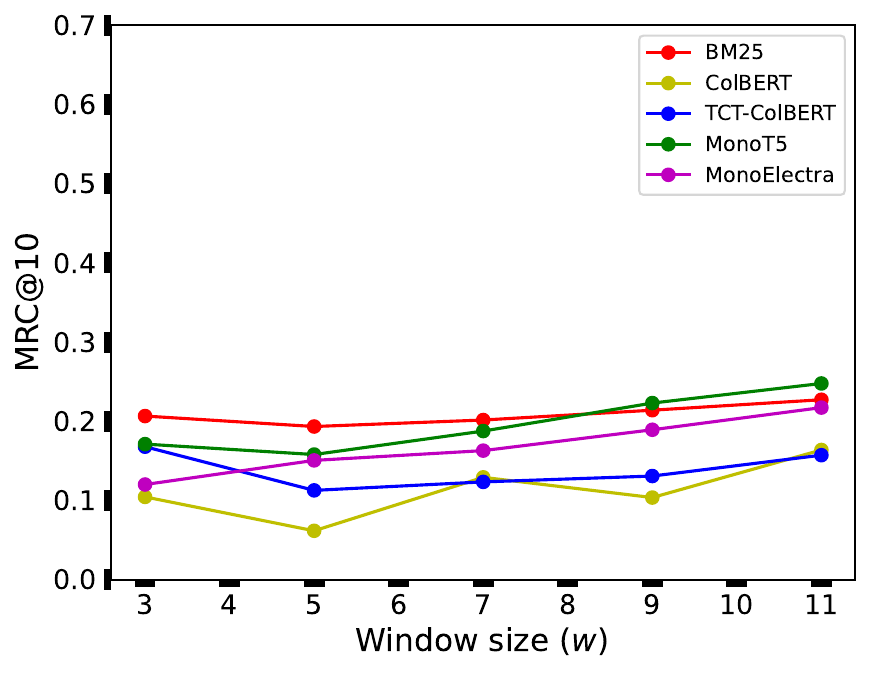}
    \caption{\small $m=1$}
    \label{fig:sub1}
\end{subfigure}
\begin{subfigure}[b]{0.32\textwidth}
    \includegraphics[width=\textwidth]{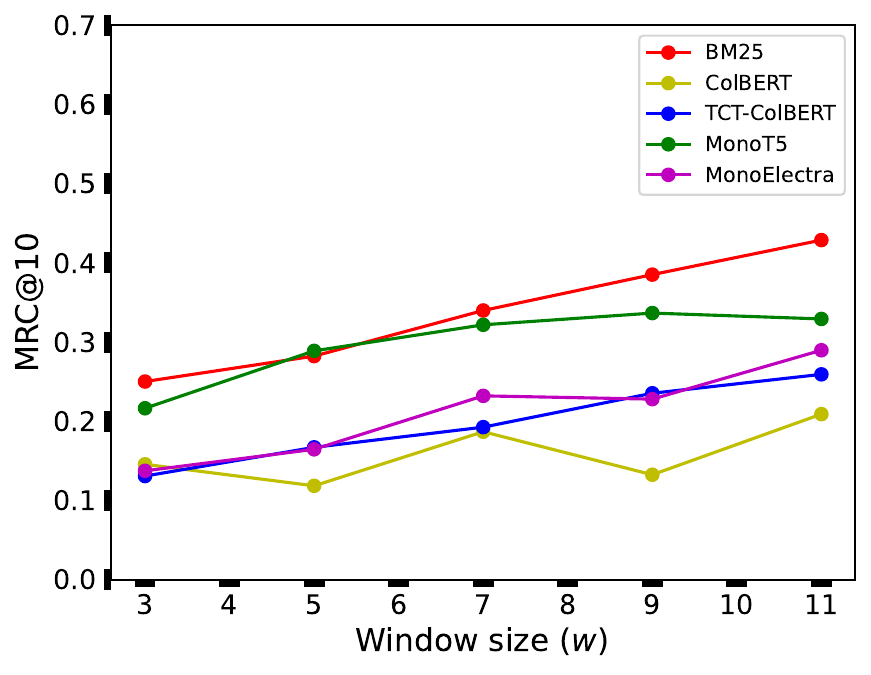}
    \caption{\small $m=3$}
    \label{fig:sub2}
\end{subfigure}
\begin{subfigure}[b]{0.32\textwidth}
    \includegraphics[width=\textwidth]{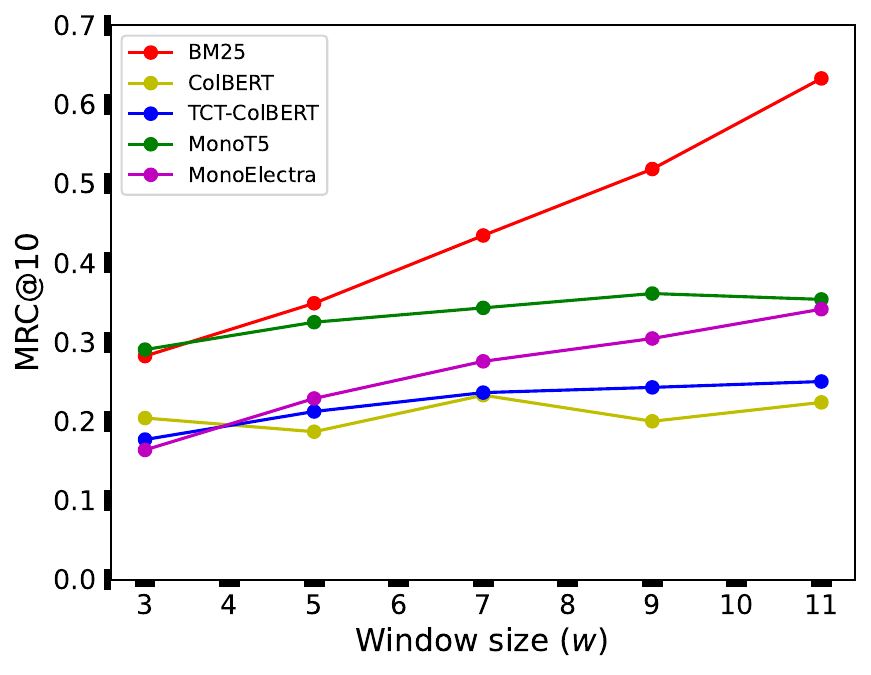}
    \caption{\small $m=5$}
    \label{fig:sub3}
\end{subfigure}
\caption{
\small
Effects of varying the size of the explanation units on the intrinsic consistency of the explanations (MRC of Equation \ref{eq:mrc}) on the top-10 search results obtained with several NRMs. A comparison across the plots reveals the the effect of the variations in the number of rationales provided as explanations ($m$). A general observation is that a higher number of explanations coupled with larger explanation units tend to provide more consistent explanations. 
}
\label{fig:passage_mrc_sensitivity}
\end{figure}

\textbf{Sentence vs. word-window rationales:}
Table \ref{tab:passageres} shows that the MRC values of the sentence level rationales are higher than that of the word window based ones, which indicates that sentence-level rationales are more consistent than word-window based ones. This can likely be attributed to the fact that, firstly, sentences retain more informative content of a document which is useful for providing consistent explanations, and that secondly, transformer-based approaches due to their underlying language models usually favour well-formed sentences~\cite{abnirml}. However, short word segments are more preferable from a user's perspective because reading them potentially requires less effort, as a  result of which existing work on text-based explanations have commonly used short word windows as rationales \cite{lime,SHAP,verma2019lirme,shap_ir}.
It is also observed that the sentence rationales of NRMs are more consistent in comparison to the BM25 ones. This is due to the fact that the linguistic coherence of word sequences in sentence-level explanations work well for the NRMs, BM25 being oblivious of word ordering.   

\textbf{Explaining top-10 vs. top-50 results:}
From Table \ref{tab:passageres}, we see that the MRC@50 values are lower than those of MRC@10, which indicates that it is easier to explain a smaller number of search results. Again, this is expected because documents towards the very top ranks would actually contain text segments that are potentially relevant to the information need thus amenable to more consistent explanations.

\begin{table}[t]
\centering
\caption{
\small
A comparison of the relevance and explanation consistency (both intrinsic and extrinsic) of different IR models on the MS-MARCO document collection for top-10 and top-50 search results with sentence-level rationales. Similar to Table \ref{tab:passageres}, BM25 results (shown in gray) are included as a point for comparison with the NRM explanation effectiveness.
The best results in terms of relevance and the explanation effectiveness of the NRMs are bold-faced.
The number of sentences used as rationales ($m$) was set to $1$ for these results.
\label{tab:docres}
}
\small
\begin{adjustbox}{width=.65\columnwidth}
\begin{tabular}{@{}l cc cc cc@{}}
\toprule
& \multicolumn{2}{c}{} & \multicolumn{4}{c}{Explanation Effectiveness} \\
\cmidrule(r){4-7}
& \multicolumn{2}{c}{Relevance (nDCG)} & \multicolumn{2}{c}
{Intrinsic (MRC)} & \multicolumn{2}{c}{Extrinsic (MER)} \\
\cmidrule(r){2-3}
\cmidrule(r){4-5}
\cmidrule(r){6-7}
Model & top-10 & top-50 & top-10 & top-50 & top-10 & top-50 \\ 
\midrule
BM25 & \cellcolor{lightgray}0.5213 & \cellcolor{lightgray}0.5325 & \cellcolor{lightgray}0.1660 & \cellcolor{lightgray}0.1798 & \cellcolor{lightgray}0.2024 & \cellcolor{lightgray}0.1823\\
\midrule
ColBERT & 0.5675 & 0.5686 & 0.2064 & 0.2801 & 0.1846 & 0.1647\\
TCT-ColBERT & 0.5787 & 0.5654 & 0.2420 & 0.2721 & \textbf{0.2069} & 0.1701\\
MonoT5 & 0.6045 & 0.6048 & 0.2133 & \textbf{0.3265} & 0.2039 & 0.1703\\
MonoElectra & \textbf{0.6185} & \textbf{0.6051} & \textbf{0.2493} & 0.2863 & 0.1959 & \textbf{0.1739}\\

\bottomrule
\end{tabular}
\end{adjustbox}
\end{table}

\para{Document Ranking Results}
Similar to Table \ref{tab:passageres}, in Table \ref{tab:docres} we show the results on the MS-MARCO document ranking task. For this task, since documents are much larger than MS-MARCO passages, we restrict the granularity of explanation units to sentences only. In addition to the intrinsic explainability measure, Table \ref{tab:docres} also reports values for the extrinsic one obtained with sub-document level relevance. Following are the key observations from Table \ref{tab:docres}.

\textbf{Intrinsic explanation consistency:}
The results are similar to that of the passage task (Table \ref{tab:passageres}) in the sense that we note considerable variations in the reported MRC values across IR models. Also, similar to the passage task, we observe that the intrinsic explanation consistency of the NRMs is better than that of BM25's.

\textbf{Extrinsic explanation consistency:}
It is observed from the MER columns of Table \ref{tab:docres} that the extrinsic evaluation effectiveness is not necessarily correlated with the intrinsic one, e.g., the best model in terms of intrinsic consistency (MRC) of explanations of top-10 documents is MonoElectra (similar to the passage ranking task), whereas the best model with the extrinsic explanation consistency is TCT-ColBERT. This indicates that the parameterised ranking function of TCT-ColBERT puts more attention, on an average, to the relevant pieces of text in comparison to the other NRMs.

In relation to RQ-3, we can thus comment that similar to the MRC variations, we do observe a noticeable variations in the extrinsic explainability measure as well indicating that some NRMs are substantially better explainable than others in terms of the relevance of the rationales.

\textbf{Explaining top-10 vs. top-50 results:}
Different from Table \ref{tab:passageres}, in Table \ref{tab:docres} we observe that the intrinsic explanations for the top-50 results are better than the top-10 ones (compare MRC@10 vs. MRC@50 across the different IR models). This can be attributed to the fact that due to the large length of the documents, it is still likely to find some partially relevant sentences even in documents that are not at the very first search result page (e.g., within top-10). An NRM by putting attention to these partially relevant sentences can yield consistent explanations even within the top-50 set of documents.

However, the trend is reversed for the extrinsic measure MER, where we observe that the top-10 results are better than top-50 ones, the reason being it is more likely for the rationales of the top-10 documents to overlap with the relevant passages. 

\para{Further Analysis}

For the passage ranking task, we now investigate the effects of the variations in i) the number of rationales used to explain the search results, i.e., $m$, and ii) the size of the explanations in terms of the number of words or sentences depending on the granularity - word windows, or sentences, i.e., $w$.
Figure \ref{fig:passage_mrc_sensitivity} shows the effect of variations in the length of the rationales (in terms of the number of words constituting the explanation units) for $3$ different values of the number of explanations provided.

We observe that, first, the explanation consistency increases with an increase in the number of rationales provided (compare the MRC values across the different plots). This is expected because with a higher number of rationales, each document's attention-focused representation (Equation \ref{eq:pseudodoc}) tends towards its original representation, i.e., the explanations themselves tend to cover the entire document. While this shows up as increased values of MRC, as pragmatic reasons, too large a value of $m$ is not desirable. 
Moreover, we also observe that for most NRMs, increasing the rationale length leads to more consistent explanations. Too large text windows can lead to inclusion of potentially non-relevant text in the explanations, which due to the semantic incoherence with the queries can lead to decreased MRC values (as can be seen from the drops in MRC from $w=9$ to $w=11$ in most cases).

\begin{figure}[t]
\centering
\begin{subfigure}[b]{0.4\textwidth}
\includegraphics[width=\columnwidth]{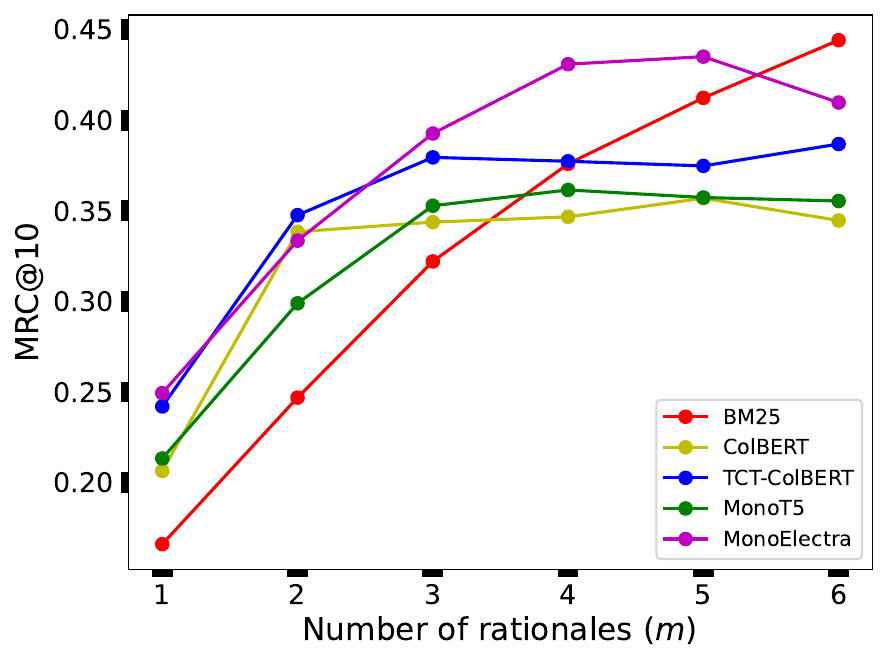}
\caption{\small
Intrinsic Consistency}
\label{fig:doc_intrinsic}
\end{subfigure}
\begin{subfigure}[b]{0.4\textwidth}
\includegraphics[width=\columnwidth]{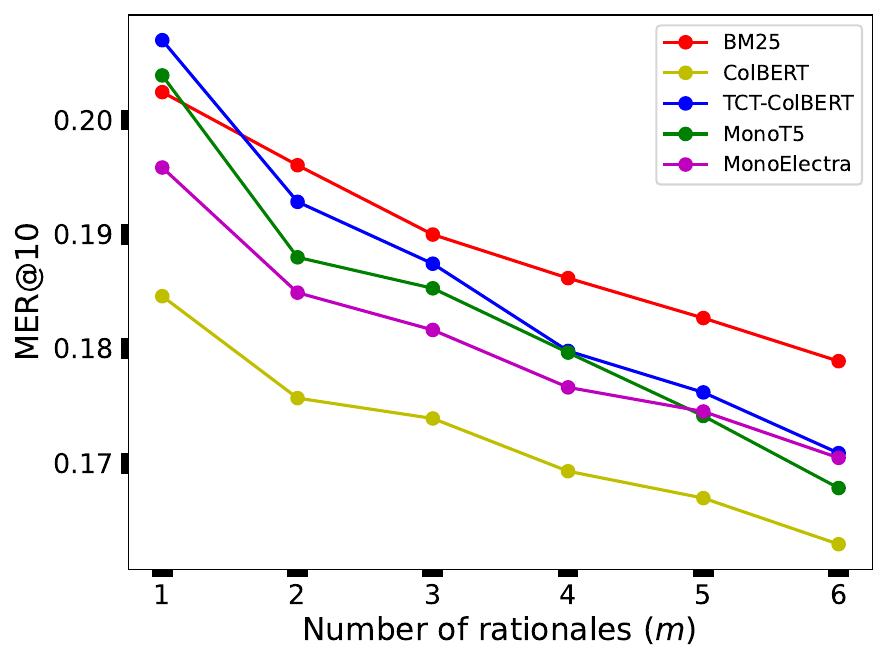}
\caption{\small
Extrinsic Consistency}
\label{fig:mer}
\end{subfigure}
\caption{
\small
Effect of the number of rationales on the explanation consistency metrics across different NRMs for the MS-MARCO document ranking task.
}
\label{fig:doc_task_parameters}
\end{figure}

\begin{figure}[t]
    \centering    
    \begin{subfigure}[b]{0.18\columnwidth}
        \includegraphics[width=\textwidth]{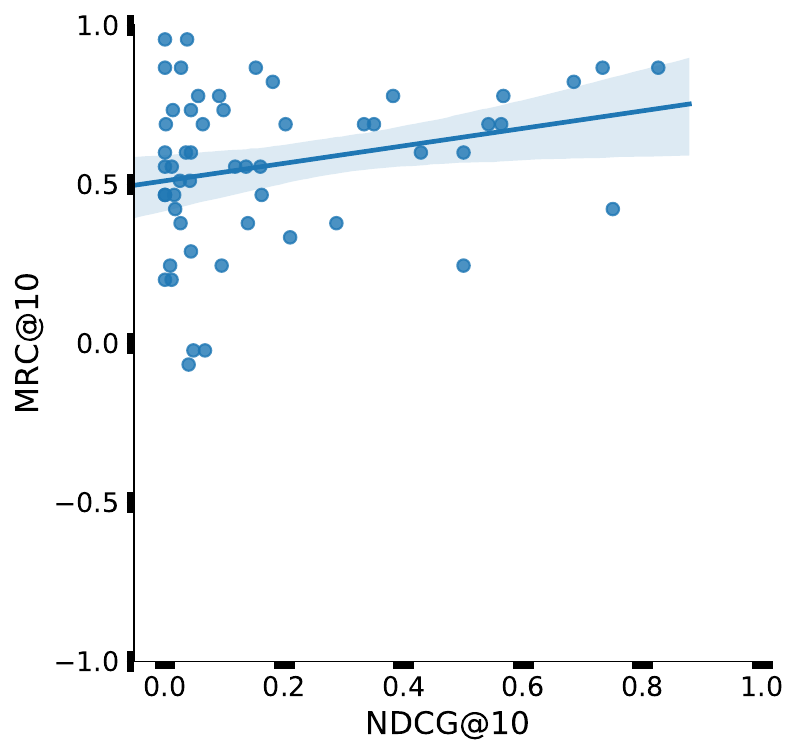}
        \caption{\scriptsize BM25}
        \label{sfig:passage-scatter-bm25}
    \end{subfigure}
    \begin{subfigure}[b]{0.18\columnwidth}
        \includegraphics[width=\textwidth]{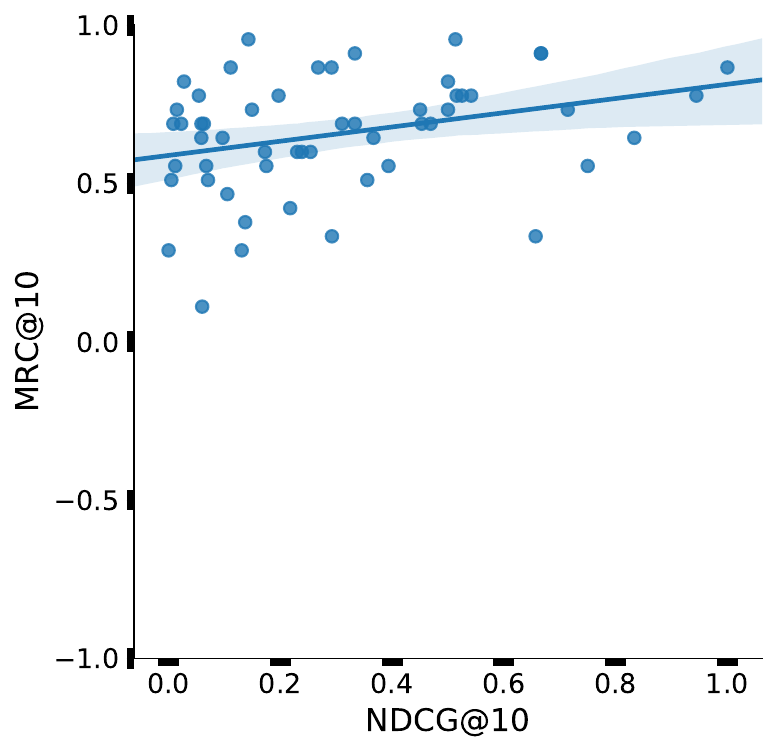}
        \caption{\scriptsize ColBERT}
        \label{sfig:passage-scatter-colbert}
    \end{subfigure}
    \begin{subfigure}[b]{0.2\columnwidth}
        \includegraphics[width=\textwidth]{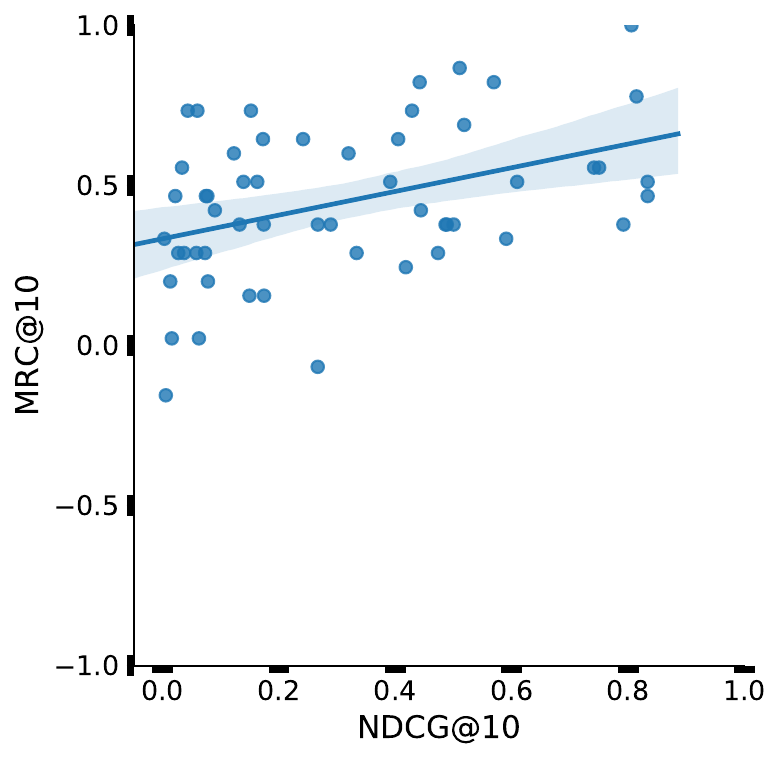}
        \caption{\scriptsize TCT-ColBERT}
        \label{sfig:passage-scatter-tct}
    \end{subfigure}
    \begin{subfigure}[b]{0.18\columnwidth}
        \includegraphics[width=\textwidth]{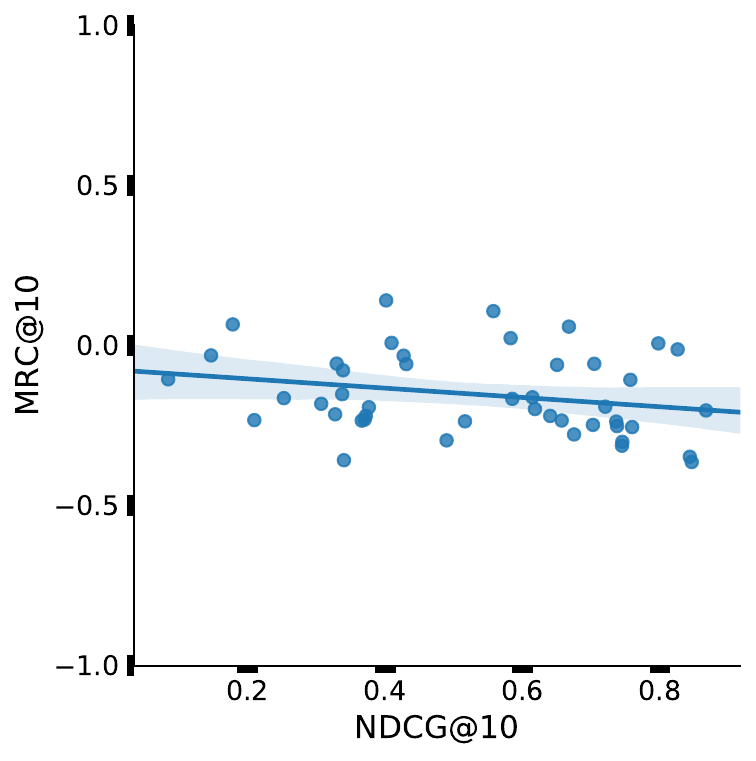}
        \caption{\scriptsize MonoT5}
        \label{sfig:passage-scatter-monot5}
    \end{subfigure}  
    \begin{subfigure}[b]{0.18\columnwidth}
        \includegraphics[width=\textwidth]{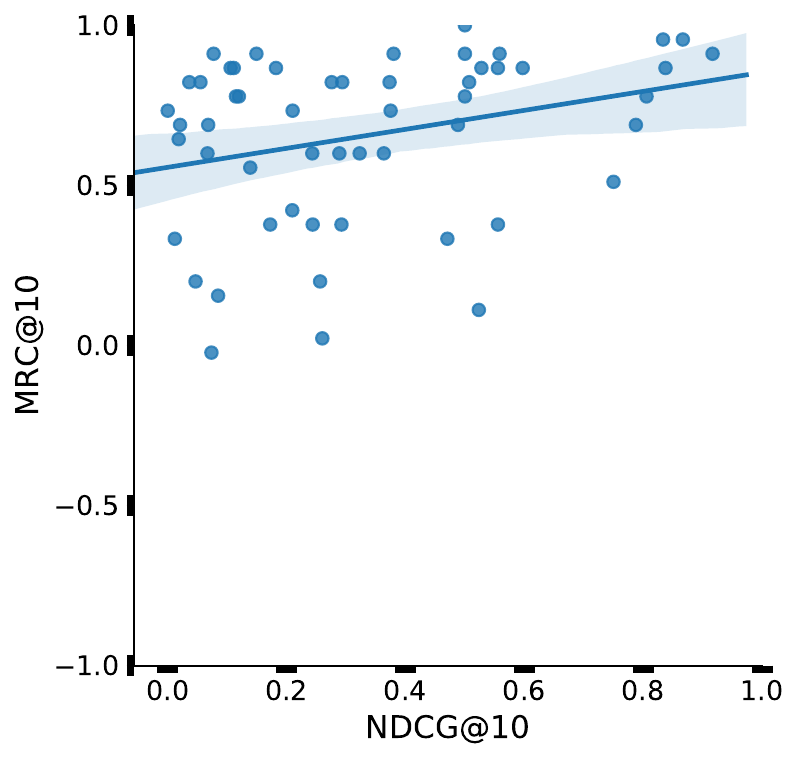}
        \caption{\scriptsize \monoelectra}
        \label{sfig:passage-scatter-electra}
    \end{subfigure}
    \caption{
    \small
    Per-query comparisons of the relevance and intrinsic explanation consistency measures (MRC) for different IR models on the MS-MARCO passage ranking task.}
    \label{fig:passage-scatter}
\end{figure}

We now conduct a similar parameter variation analysis for the document ranking task. Since the explanations are at sentence-level, this task does not involve the parameter $w$ (the size of explanations). Figures \ref{fig:doc_intrinsic} and \ref{fig:mer} show the effects of variations of the number of rationales on the intrinsic and the extrinsic explanation consistencies, respectively.
While with an increase in $m$, the intrinsic measure mostly increases, a reverse trend is observed for the sub-document relevance-based extrinsic approach. The argument on why MRC increases with an increase in the number of rationales is the same as that of passages, i.e., the attention-focused representation of a document tends to better represent its core topic.
On the other hand, a likely reason for the decrease in the MER values is that the sentences with lower explanation weights (but still included as a part of the explanations due to relatively higher values of $m$) represent segments of a document that are potentially not relevant to the query. This means that the proportion of sub-document level relevant information in the explanations (Equation \ref{eq:mer}) progressively decreases.






\begin{figure}[t]
    \centering    
    \begin{subfigure}[b]{0.18\columnwidth}
        \includegraphics[width=\textwidth]{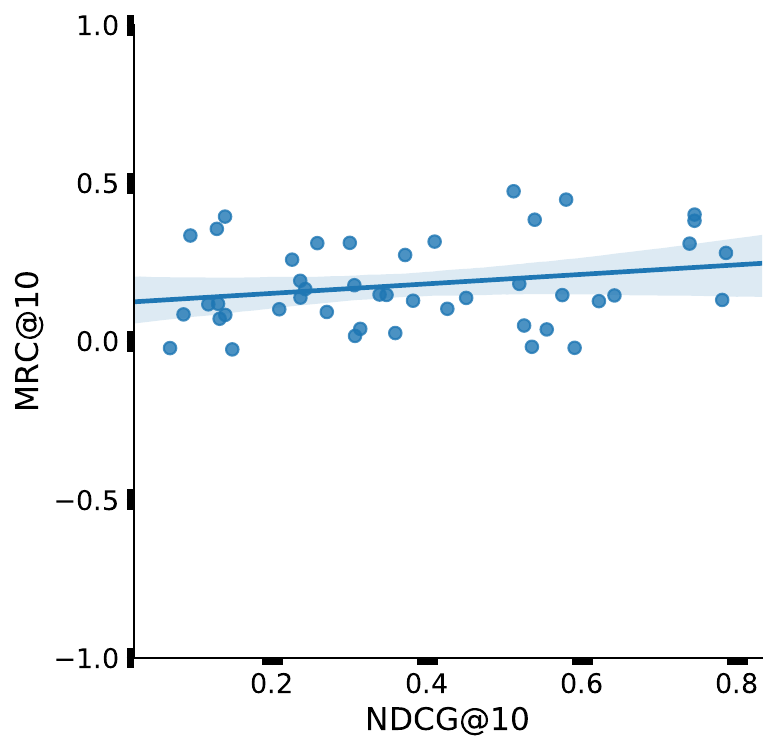}
        \caption{\scriptsize BM25}
        \label{sfig:doc-scatter-mrc-bm25}
    \end{subfigure}
    \begin{subfigure}[b]{0.18\columnwidth}
        \includegraphics[width=\textwidth]{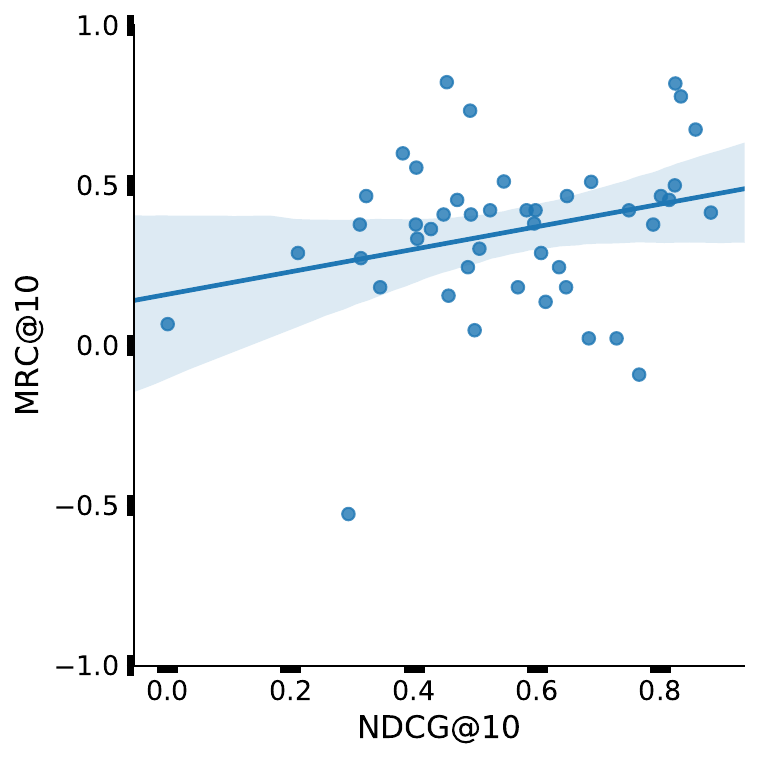}
        \caption{\scriptsize ColBERT}
        \label{sfig:doc-scatter-mrc-colbert}
    \end{subfigure}
    \begin{subfigure}[b]{0.2\columnwidth}
        \includegraphics[width=\textwidth]{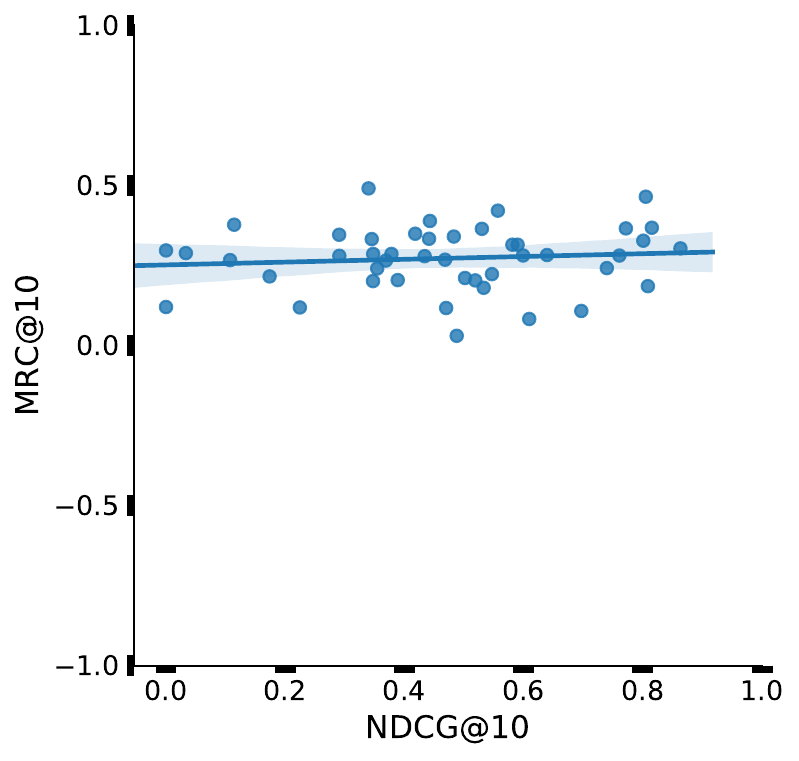}
        \caption{\scriptsize TCT-ColBERT}
        \label{sfig:doc-scatter-mrc-tct}
    \end{subfigure}
    \begin{subfigure}[b]{0.18\columnwidth}
        \includegraphics[width=\textwidth]{figs/doc_corr/monoT5.pdf}
        \caption{\scriptsize MonoT5}
        \label{sfig:doc-scatter-mrc-monot5}
    \end{subfigure}  
    \begin{subfigure}[b]{0.18\columnwidth}
        \includegraphics[width=\textwidth]{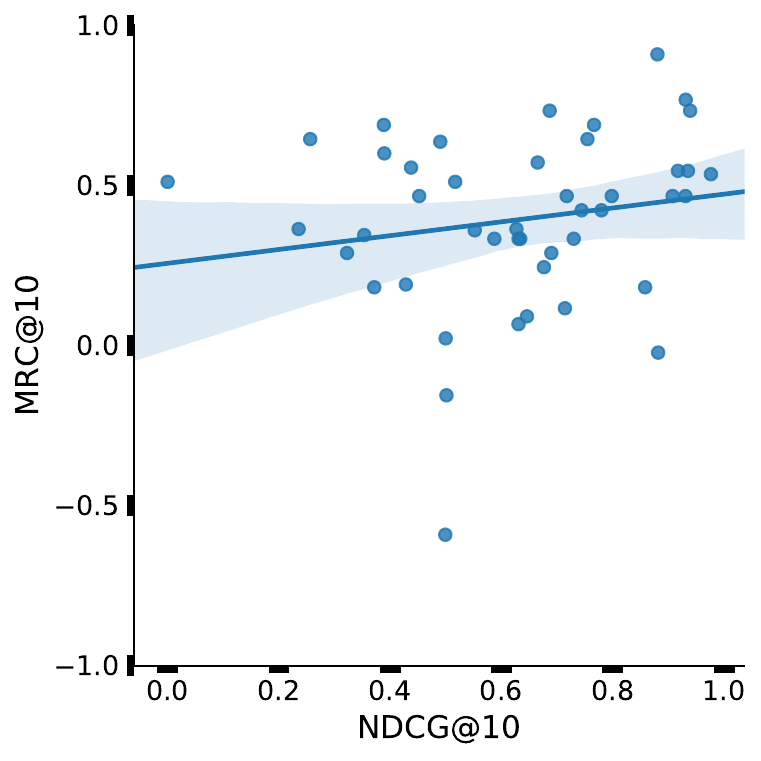}
        \caption{\scriptsize MonoElectra}
        \label{sfig:doc-scatter-mrc-electra}
    \end{subfigure}
    \caption{
    \small
Per-query comparisons of the relevance and intrinsic explanation consistency measures (MRC) for different IR models on the MS-MARCO document ranking task.}
    \label{fig:doc-scatter-mrc}
\end{figure}

\begin{figure}[t]
    \centering
    \begin{subfigure}[b]{0.18\columnwidth}
        \includegraphics[width=\textwidth]{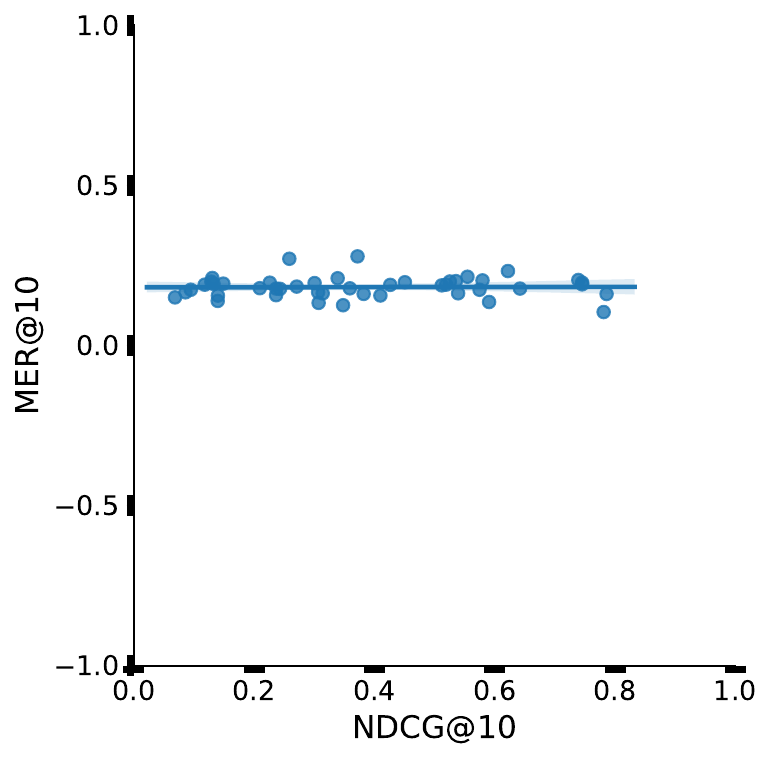}
        \caption{\scriptsize BM25}
        \label{sfig:doc-scatter-mer-bm25}
    \end{subfigure}
    \begin{subfigure}[b]{0.18\columnwidth}
        \includegraphics[width=\textwidth]{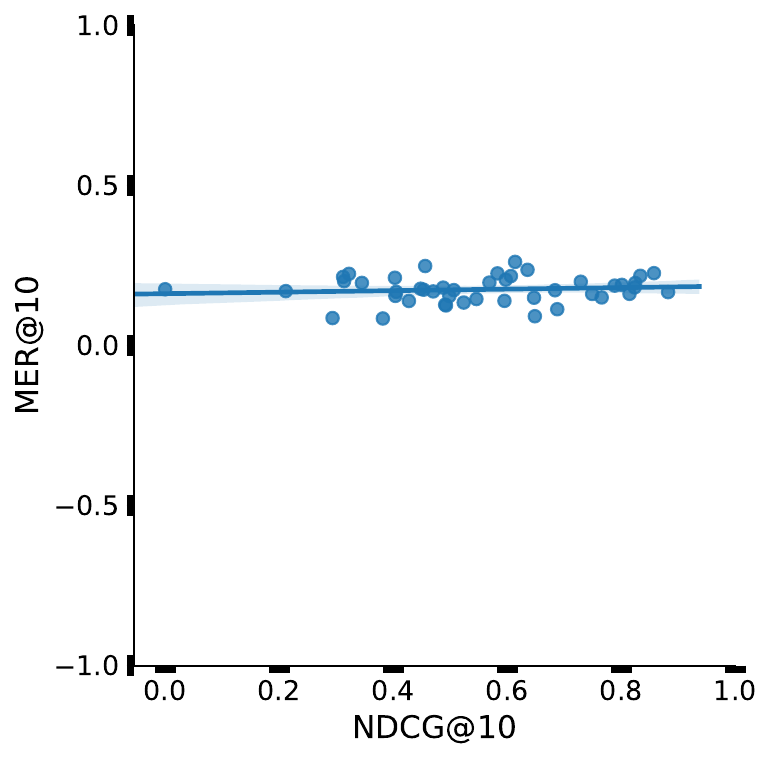}
        \caption{\scriptsize ColBERT}
        \label{sfig:doc-scatter-mer-colbert}
    \end{subfigure}
    \begin{subfigure}[b]{0.2\columnwidth}
        \includegraphics[width=\textwidth]{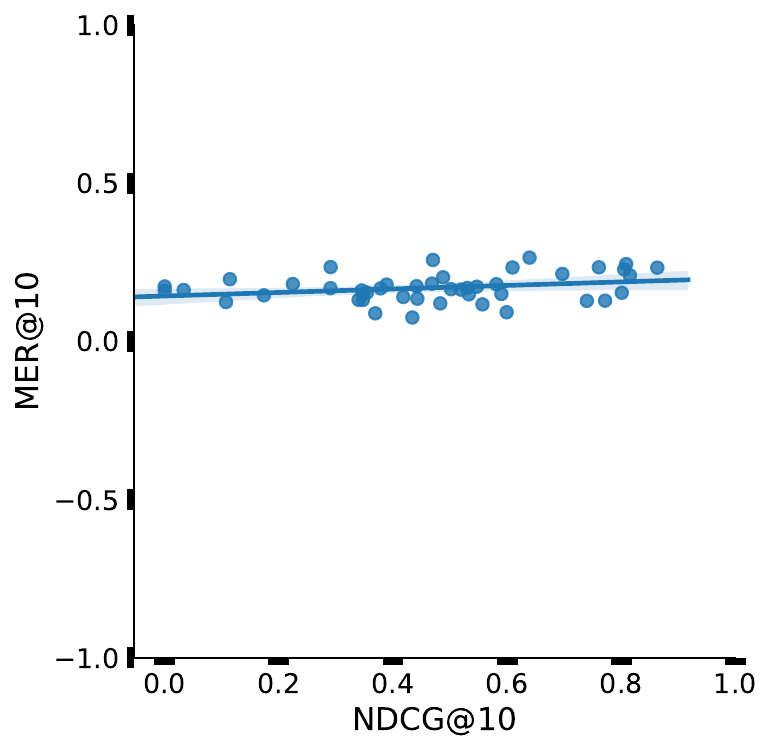}
        \caption{\scriptsize TCT-ColBERT}
        \label{sfig:doc-scatter-mer-}
    \end{subfigure}
    \begin{subfigure}[b]{0.18\columnwidth}
        \includegraphics[width=\textwidth]{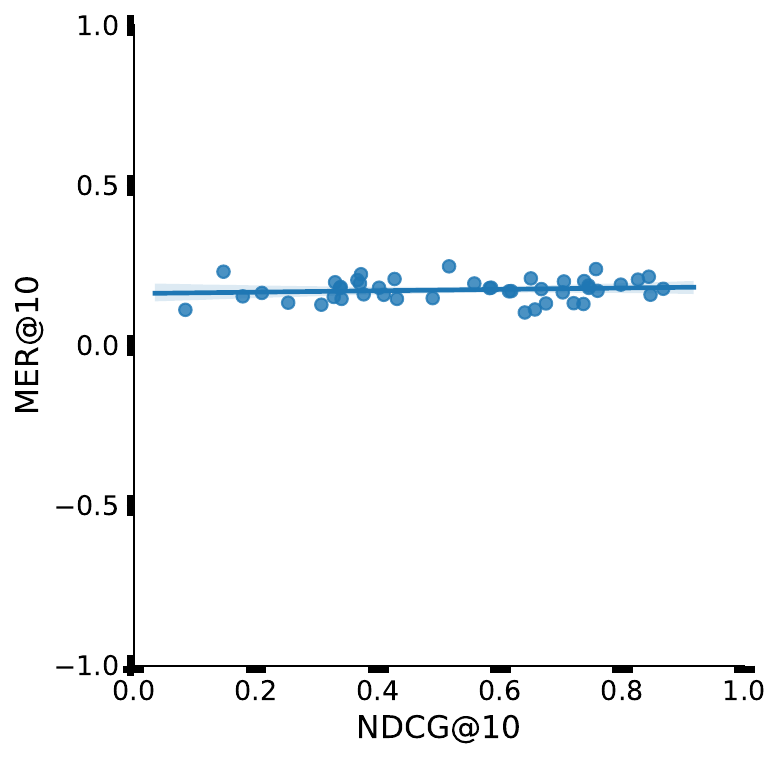}
        \caption{\scriptsize MonoT5}
        \label{sfig:doc-scatter-mer-monot5}
    \end{subfigure}  
    \begin{subfigure}[b]{0.18\columnwidth}
        \includegraphics[width=\textwidth]{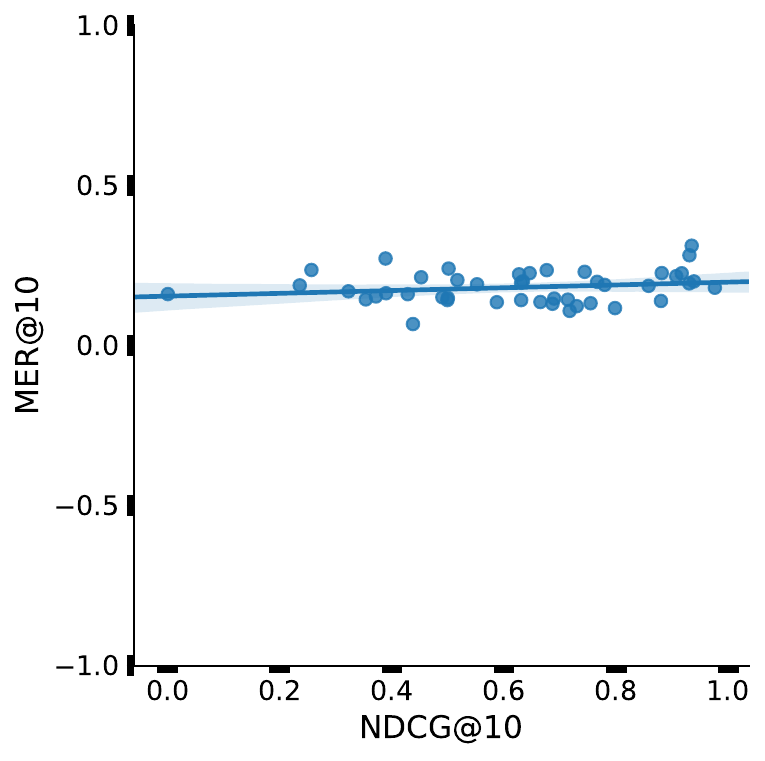}
        \caption{\scriptsize MonoElectra}
        \label{sfig:doc-scatter-mer-electra}
    \end{subfigure}
\caption{
\small
Per-query comparisons of the relevance and extrinsic explanation consistency measures (MER) for different IR models on the MS-MARCO document ranking task.
}
\label{fig:doc-scatter-mer}
\end{figure}

\para{Explanation Effectiveness vs. Relevance}

In relation to research questions RQ-2 and RQ-4, we now investigate the correlation between a relevance measure (specifically, nDCG@10 as used in our experiments) and the explanation consistency measures. Tables \ref{tab:passageres} and \ref{tab:docres} already demonstrated that the best performing NRM in terms of relevance is not necessarily the one yielding the most consistent explanations.
This observation is reinforced by Figure \ref{fig:passage-scatter} which shows a scatter-plot of the per-query nDCG@10 vs. the MRC values.
An interesting observation from Figure \ref{fig:passage-scatter} is that all the models (even BM25) registers only a small correlation between the relevance and explanation consistency. MonoT5 even yields a negative correlation of MRC with nDCG@10. This suggests that explanation consistency can potentially serve as an evaluation dimension complementary to that of relevance.

Similar observations can also be made from Figures \ref{fig:doc-scatter-mrc} and \ref{fig:doc-scatter-mer}. MER is slightly more correlated with relevance because it involves measuring an overlap of the rationales with the relevant segments of a document. However, the insights gained from the MER metric is still different from both relevance and the intrinsic consistency, which, in turn, suggests that MER like MRC can potentially be used as another dimension to evaluate NRMs.


\begin{table*}[t]
\centering
\small
\caption{\small The top-2 documents of the best performing query in terms of the intrinsic consistency measure for MonoElectra on the MS-MARCO passage ranking task. The top-3 rationales are shown with 3 different color shades (a deeper shade indicating a higher attention weight).
\label{tab:examples}
}
\begin{adjustbox}{width=0.7\textwidth}
\begin{tabularx}{\textwidth}{@{}lX@{}}
\toprule
Query & Rationales of the top-$3$ documents \\
\midrule
\multirow{3}{*}{What is the UN FAO?}
&  United Nations Food and Agriculture Organization (FAO) The \mybox[fill=red!30]{Food and Agriculture Organization of} the United Nations is an agency \mybox[fill=orange!30]{that leads international efforts} to defeat hunger. The \mybox[fill=red!30]{Food and Agriculture Organization of} the United Nations is an agency of the United Nations that leads international efforts to defeat hunger.
\\
\cmidrule{2-2}
(MRC@10 = 0.9112) &
Role of the FAO. \mybox[fill=yellow!30]{The Food and Agriculture Organization} \mybox[fill=orange!30]{of the United Nations is} \mybox[fill=red!30]{an agency of the United} Nations that leads international efforts to defeat hunger. Serving both developed and developing countries, FAO acts as a neutral forum where all nations meet as equals to negotiate agreements and debate policy.
\\
\bottomrule
\end{tabularx}
\end{adjustbox}
\end{table*}

\para{Sample Explanations}
Table \ref{tab:examples} shows an example query with the best intrinsic consistency measure on the MS-MARCO passage ranking task. It can be seen that rationales provided for this query does also qualitatively indicate pertinent explanations. This can be seen from the fact that the NRM (MonoElectra) has been able to bridge the semantic gap between the query term `FAO', and the ones in the top-ranked documents such as `Food', `Agricultural' and `Organization'. The same argument also applies for the query term `UN', and the document terms `United' and `Nations'.


\para{Concluding Remarks}

We introduced a evaluation framework wherein rationales -- represented as text snippets from the document -- are available alongside search results. These rationales can then be evaluated alongside traditional relevance measures to assess the interpretability of the models.
Using a basic occlusion-based technique for producing rationales, we find that the \textbf{systems that produce the most relevant results may not necessarily be the most explainable} -- both in terms of intrinsic and extrinsic measures of explainability.

As a future work, tools could be developed to enable easier qualitative evaluation of rationales~\cite{jose:sigir2021-diffir}, and user studies could be conducted to test how helpful rationales are to end-users.


\bibliographystyle{splncs04}
\bibliography{ref,nrms,trust}

\end{document}